\documentclass[twocolappendix,apj]{emulateapj}

\usepackage{perpage}
\usepackage{hyperref}
\usepackage{subdepth}
\usepackage{pdftexcmds}

\def\I{\,\textsc{i}}
\def\II{\,\textsc{ii}}

\def\IV{\,\textsc{iv}}

\shorttitle{Variability in Near-IR Synchrotron From Cas A}
\shortauthors{Kilpatrick, Rieke, \& Eriksen}

\begin{document}


\title{Variability in the Near-Infrared Synchrotron Emission From Cassiopeia A}

\author{Charles D. Kilpatrick$^1$, George H. Rieke$^1$, \& Kristoffer Eriksen$^2$}
\affil{$^1$Steward Observatory, University of Arizona, Tucson, AZ 85721\\
	$^2$XTD-IDA, Los Alamos National Laboratory, Los Alamos, NM 87545}


\begin{abstract}
	
We present multi-epoch $K_{s}$ band imaging of the supernova remnant Cassiopeia A (Cas A). The morphology of the emission in this band is generally diffuse and filamentary, consistent with synchrotron radiation observed at radio wavelengths.  However, in one region to the southwest of the remnant, compact knots of emission appear to be entrained in the ejecta and have the same proper motion as ejecta observed at similar projected radii.  The presence of these knots suggests that material with high magnetic field strength contributes significantly to synchrotron emission at these wavelengths.  We analyze these knots at $J$, $H$, and $K_{s}$ bands as well as in $3.5-8~\text{micron}$ emission and at $6~\text{cm}$ where synchrotron emission is dominant and we find that the $K_{s}$ band emission falls along the expected synchrotron spectrum. Using multi-epoch data, we calculate the magnetic field strength and electron density for a population of near-infrared synchrotron-emitting electrons. We find electron densities from $1,000-15,000~\text{cm}^{-3}$ and magnetic field strengths from $1.3-5.8~\text{mG}$. These magnetic field strengths are an order of magnitude higher than inferred from the much lower angular resolution gamma-ray observations toward Cas A.  We conclude that dense knots of post-shock material behind the Cas A shock front are emitting synchrotron emission in a compressed and enhanced magnetic field.

\end{abstract}




\section{Introduction}

Acceleration of cosmic rays (CRs) in supernova remnants (SNRs) contributes significantly to the population of high energy electrons up to the ``knee'' of the CR spectrum around $10^{15}~\mathrm{eV}$.  This population has been observed toward SNRs via synchrotron radiation and other nonthermal emission from radio through gamma ray energies. The diffusive shock acceleration model has largely succeeded in tying the radio synchrotron spectrum observed toward SNRs to GeV electrons \citep{bell78}. At higher energies, electrons with $E > 1~\text{TeV}$ have been associated with X-ray emission in the form of synchrotron radiation \citep{reynolds98}, inverse Compton scattering \citep{tanimori+98,porter+06}, and nonthermal brehmsstralung radiation for the hardest X-rays above $100~\text{keV}$ \citep{vink+08}. GeV gamma-rays from SNRs likely originate from leptonic CRs, especially toward SNRs where particle acceleration is enhanced due to interaction with a molecular cloud \cite[e.g.,][]{castro+13}. Even for gamma-rays with $E \geq 1~\text{TeV}$, it is ambiguous to what extent the emission originates from leptonic as opposed to hadronic particle acceleration \citep{ellison+10,inoue+12,slane+15}.

Near-infrared $K_{s}$ band emission from SNRs is thought to be dominated by synchrotron emission.  At these energies, synchrotron radiation requires electrons with higher energies ($\sim 0.2~\mathrm{TeV}$) than those that emit predominantly in the radio.  In support of this hypothesis, spectroscopic measurements of the $K_{s}$ band spectral index $\alpha$ (where $F_{\nu} \sim \nu^{\alpha}$) toward Cassiopeia A (Cas A) with $-0.80 < \alpha < -0.67$ are consistent with a synchrotron spectrum \citep{wright+99,rho+03,eriksen+09}.  In addition, the fractional polarization of $K_{s}$ band emission (5-10\%) is consistent with measurements of synchrotron emission around $6~\text{cm}$ \citep{jones+03}.  Measurements of the emission in this region therefore provide a unique probe of the synchrotron spectrum while offering a test against which radio, X-ray, and gamma-ray synchrotron measurements can be compared.

One of the most significant advantages of near-infrared measurements of Cas A involves the timescales of synchrotron losses.  Given the magnetic field strengths and electron energies involved, infrared-emitting electrons are likely to have been accelerated no more than $\sim80$ yr ago while shocks can produce infrared-emitting electrons on timescales of $\sim1$ yr \citep{jones+03,rho+03}. Therefore, multi-epoch measurements of near-infrared synchrotron emitting material on timescales of several years are likely to be sensitive to variability in the acceleration of electrons over baselines of $1-10~\text{yrs}$.  In turn, this variability is a direct probe of the electron density and magnetic field accelerating these electrons.

Additionally, current magnetic field strength estimates for the radio-emitting plasma in Cas A are based on gamma-ray fluxes measured with poor angular resolution and thus averaged over multiple acceleration sites. The magnetic field strength over this region is related to the distribution of relativistic particle energies and the bremsstrahlung flux emitted by those particles, which is thought to be the dominant emission process at GeV energies.  Cas A is detected by \textit{Fermi} with a beam size of $0^{\circ}.1$ as a single GeV source and this emission is well-fit by a leptonic model with a magnetic field strength of $B\approx0.12~\text{mG}$ \citep{abdo+10}.  Alternative analyses based on radio, infrared, X-ray, and \textit{Fermi} gamma-ray data suggest $B\approx0.05-0.30~\text{mG}$ \citep{araya+10}, $0.23-0.51~\text{mG}$ \citep{saha+14}, and $0.2-0.4~\text{mG}$ \citep{zirakashvili+14}.  Although the production of GeV gamma-ray emission via the leptonic process predicts short cooling timescales and thus emission originating near the Cas A forward shock \citep[e.g.,][]{esposito+96,gotthelf+01,abdo+10}, these measurements cannot be localized to specific regions where significant magnetic field amplification might occur.

In all of these studies, in effect the magnetic field strengths are averaged over the entire relativistic electron population of Cas A. On small scales, compression and turbulent amplification may lead to significant magnetic field amplification.  This amplification should be measurable in the near-infrared when nonthermal emission is observed over multiple epochs.  In this paper we use multi-epoch $K_{s}$ band imaging to constrain the proper motions and variability in knots of synchrotron emission toward Cas A.  We look for any features that are well-resolved in the imaging and compare them to archival data in other wavelengths to verify that the emission from these features falls along a synchrotron spectrum.  Finally, we derive magnetic field strengths and electron densities for these features and compare them to values calculated over the entire remnant.

\section{Observations}

We obtained near-infrared $K_{s}$ band imaging of Cas A using PISCES on the Bok 2.3m telescope on 11 Nov. 2013.  The PISCES wide-field camera \citep{mccarthy+01} has a field of view of 8\arcmin.5 and a pixel size of roughly 0\arcsec.5 and we were able to observe the entire SNR in a single pointing.  We employed a $K_{s}$ band filter centered at $2.22~\text{microns}$ with a width of approximately $0.51~\text{microns}$.

We used 12 s individual exposures and alternated sky exposures in an on (source)-off-off-on pattern.  For each individual exposure, we added a random 30\arcsec ``wobble'' vector in order to observe the source at a random position on the array.  The total on-source exposure time in $K_{s}$ band was roughly 32 minutes.  We had 1\arcsec.2 seeing for the entirety of our observation.

Standard image reductions were performed using IRAF\footnote{IRAF, the Image Reduction and Analysis Facility, is distributed by the National Optical Astronomy Observatory, which is operated by the Association of Universities for Research in Astronomy (AURA) under cooperative agreement with the National Science Foundation (NSF).} including bias and dark subtraction, bad-pixel removal, flat-fielding, sky subtraction, distortion correction, image stacking, and registration. Flux calibration was achieved using 2MASS \citep{skrutskie+06} stars in the same field as Cas A.  We correct for extinction toward the remnant assuming $A_{V} = 6.2$ \citep[as in][]{eriksen+09}.

\begin{figure}
		\begin{minipage}[t][][b]{0.5\textwidth}
			\includegraphics[width=\textwidth]{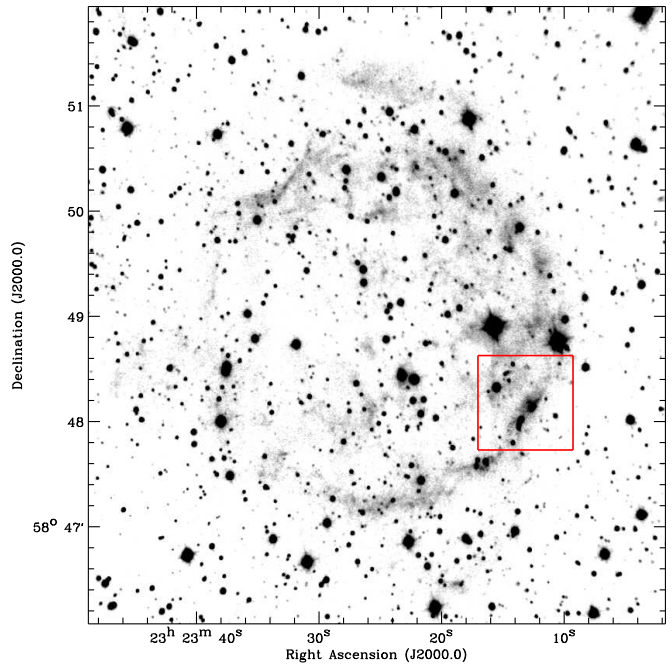}
		\end{minipage}
		\caption{Bok 2.3m PISCES $K_{s}$ band image of Cas A from 11 Nov. 2013 showing smooth synchrotron continuum emission.  Individual knots of $K_{s}$ band emission are resolved to the southwest of the remnant.  We indicate this region as depicted in \autoref{fig:knots} and \autoref{fig:radio-comp} with a red rectangle.}\label{fig:casA-K-2013}
\end{figure}

We present the final $K_{s}$ band image in \autoref{fig:casA-K-2013}.  In the subsequent analysis, we employ additional $K_{s}$ band imaging from the literature, including epochs from 2002 \citep{rho+03} and 2003 \citep{eriksen+09}.

\section{Results and Analysis}

\subsection{Fast-Moving Features}

\begin{figure*}
	\begin{center}
			\includegraphics[width=\textwidth]{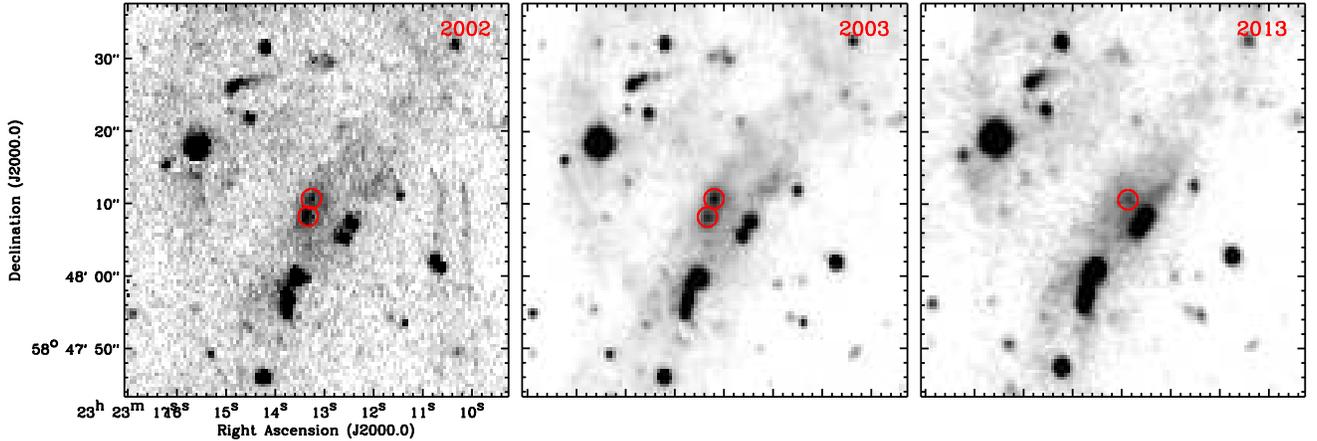}
	\caption{\scriptsize Imaging of the two synchrotron-emitting knots (SEKs) from 2002 (left), 2003 (right), and 2013 (right).  The northwestern SEK appears as extended emission near two reference stars whereas the southeastern SEK has either disappeared or moved behind the stars.  The knots are indicated with a red $1\arcsec.5$ circle in each image.}\label{fig:knots}
	\end{center}
\end{figure*}

The diffuse and filamentary structure at near-infrared wavelengths in \autoref{fig:casA-K-2013} highlights the presence of synchrotron continuum.  However, there is at least one location where the $K_{s}$ emission appears as knots of emission (\autoref{fig:knots}).  These knots appear to be entrained in the ejecta and with proper motions of at least $0\arcsec.34~\text{yr}^{-1}$, although the second knot is below the level of detectability or obscured by a star approximately $4~\text{mag}$ brighter than the knots in the final epoch.  At the distance of Cas A \citep[$3.4~\text{kpc}$][]{fesen+06}, this proper motion corresponds to a velocity of $\sim5500~\text{km s}^{-1}$, indicating that these knots of emission are likely associated with fast-moving knots \citep[FMKs; e.g.,][]{vandenbergh+70} in the ejecta.  In the following sections, we consider emission processes in $K_{s}$ band that could account for compact knots of emission entrained in the ejecta.

\subsection{Are the $K_{s}$ band knots dominated by synchrotron emission?}

A central question to this study is whether the observed $K_{s}$ band knots represent synchrotron continuum, as has been argued in the past for the underlying emission, or are they dominated by line emission from the ejecta?  Near-infrared spectroscopy of FMKs near the infrared knots reveals there is very little line emission in this waveband, with only some contribution from Br$\gamma$, He\I, [Fe\II], and [Si\IV] \citep{gerardy+01}. The bright [Si\IV] line at $1.965~\mu\text{m}$ is outside the spectral range defined by our $K_{s}$ band filter and would not contribute to the emission observed in our images.  The lack of any other strong source of line emission in typical $K_{s}$ band spectra of FMKs toward Cas A implies that the photometry is dominated by continuum emission, that is the nonthermal synchrotron spectrum.

Line emission may dominate in $J$ and $H$ bands.  In the 2003 epoch, knots 1 and 2 exhibit $m_{J} = 15.81\pm0.06$ and $15.49\pm0.04$ and $m_{H} = 15.66\pm0.07$ and $15.41\pm0.06$, respectively.  The ratio between these near-infrared bands is typical given that most near-infrared spectra toward Cas A reveal that $J$ band is strongly dominated by forbidden line emission, especially from [S\II] and [P\II] emission.  Indeed, \citet{gerardy+01} find that the flux ($F = \int F_{\nu} d\nu$) from $J$ band line emission is at least an order of magnitude greater than line emission observed in $H$ band.  However, our measurements suggest that $F = \nu F_{\nu}$ is about $F_{J} = 1840, 2480$ and $F_{H} = 1010, 1260$ (in units of $10^{-15}~\text{erg s}^{-1}~\text{cm}^{-2}$), which is atypical for FMKs.  Perhaps some other source of line emission, such as the [Fe\II] lines typically observed in slow-moving quasi-stationary floculi, can account for the unusual ratio $J$ to $H$ band ratio for knots 1 and 2, or else the nonthermal continuum must account for a significant enhancement in $H$ band.

Comparison to radio continuum suggests that the $K_{s}$ band is morphologically similar to wavebands dominated by synchrotron emission.  \citet{gerardy+01} found that the $K_{s}$ band images have no clear optical, X-ray, or mid-infrared counterpart and the diffuse emission in these bands is most similar to radio continuum images.  \citet{jones+03} made a similar argument based on predictions of the polarization angle of synchrotron radiation from 2.2$~\mu\text{m}$, which closely matches the polarization angle observed at $6~\text{cm}$.    

\begin{figure*}
	\begin{center}
		\includegraphics[width=\textwidth]{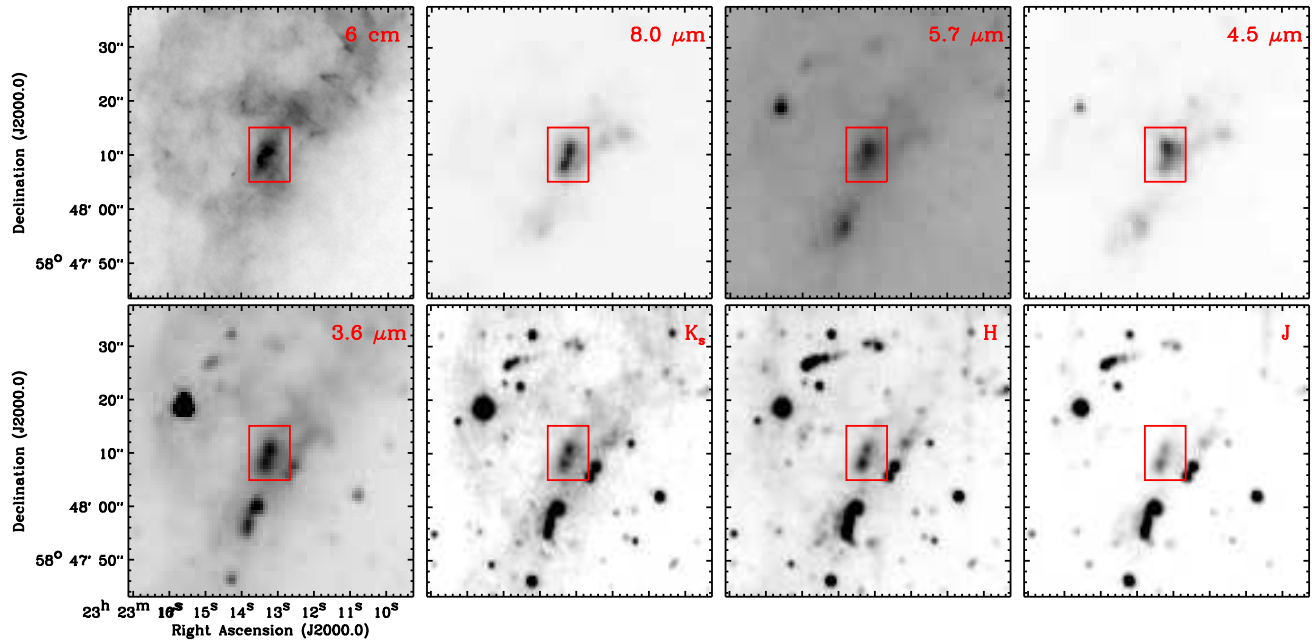}
	\caption{\scriptsize Images of the synchrotron-emitting knots at (top; left to right) 6 cm, IRAC Band 4, IRAC Band 3, IRAC Band 2, (bottom; left to right) IRAC Band 1, $K_{s}$, $H$, and $J$ bands. The red rectangle indicates the approximate position of the SEKs.}\label{fig:radio-comp}
	\end{center}
\end{figure*}

Perhaps the most convincing argument that the $K_{s}$ band features to the southwest of Cas A are dominated by synchrotron emission is the comparison between the $K_{s}$ band spectral indices and those observed in the radio.  \citet{rho+03} performed ``spectral tomography'' by matching the $K_{s}$ band brightness to radio emission across Cas A and subtracting the $K_{s}$ band emission in proportion to $(\nu_{\text{radio}}/\nu_{\text{IR}})^{\alpha}$.  This method simultaneously provides a check against the hypothesis that the near-infrared emission is well-fit by synchrotron continuum as observed in radio emission and a way to measure spectral indices across the remnant.  We performed the same analysis using archival $6~\text{cm}$ VLA imaging toward Cas A obtained from 2000-2001 \citep[see][]{roy+09}.  As we demonstrate in \autoref{fig:radio-comp}, the $K_{s}$ band knots correspond to the position of a local enhancement in the radio synchrotron emission.

At $6~\text{GHz}$, knot 1 and knot 2 have flux densities of $0.55\pm0.02$ and $0.53\pm0.02~\text{Jy}$, respectively (see \autoref{tab:1}). The radio spectral index is $\sim-0.76$ \citep{onic+15}, so extrapolating to $2.2~\mu\text{m}$ predicts flux densities of $0.27$ and $0.26~\text{mJy}$ respectively, close to the observed values of $0.37$ and $0.35~\text{mJy}$ (taking the averages for 2002 and 2003 - see \autoref{tab:1}). It is therefore plausible that the $K_{s}$ band emission is dominated by synchrotron radiation. Indeed, there are indications of a flattening of the synchrotron spectrum toward high radio frequencies \citep[][see also \citet{rho+03}]{onic+15}, which would make the extrapolation from $6~\text{GHz}$ to $2.2~\mu\text{m}$ even closer to the measured values and suggest the $K_{s}$ band emission is almost entirely due to synchrotron radiation.

\begin{deluxetable*}
{ccccccccccc}
\tabletypesize{\scriptsize}
\tablecaption{Photometry of Synchrotron-Emitting Knots Toward Cas A\label{tab:1}}
\tablewidth{0pt}
\tablehead{
Knot (\#) & 6 cm & 8.0$~\mu\text{m}$ & 5.7$~\mu\text{m}$ & 4.5$~\mu\text{m}$ & 3.6$~\mu\text{m}$ & $K_{s}$ & $K_{s}$ & $K_{s}$ & $H$ & $J$ \\
& (MJD=51794) & (53388) & (53388) & (53388) & (53388) & (52289) & (52833) & (56607) & (52833) & (52833) \\
& (mJy) & (mJy) & (mJy) & (mJy) & (mJy) & (mJy) & (mJy) & (mJy) & (mJy) & (mJy) \\
}
\startdata
1 & 551$\pm$20 & 54.1$\pm$0.1 & 1.15$\pm$0.03 & 7.94$\pm$0.04 & 0.53$\pm$0.02 & 0.37$\pm$0.06 & 0.37$\pm$0.04 & 0.31$\pm$0.04 & 0.56$\pm$0.04 & 0.76$\pm$0.04\\
2 & 530$\pm$20 & 52.7$\pm$0.1 & 1.15$\pm$0.03 & 5.37$\pm$0.04 & 0.41$\pm$0.02 & 0.39$\pm$0.06 & 0.31$\pm$0.04 & -- 	          & 0.70$\pm$0.04 & 1.02$\pm$0.04
\enddata
\tablecomments{Flux densities for knots 1 and 2 from VLA 6 cm, IRAC Band 4-1 (8.0, 5.7, 4.5, 3.6$~\mu\text{m}$), and $K_{s}$ (2002, 2003, 2013), $H$, and $J$ band emission.  We indicate the approximate Modified Julian Date (MJD) of each observation below each band.  All flux densities are in terms of mJy.  For the 2002, 2003, and 2013 $K_{s}$ band epochs, the flux densities for SEK 1 correspond to magnitudes of 15.63, 15.64, and 15.82, respectively, while those for the 2002 and 2003 epochs for SEK 2 are 15.59 and 15.82, respectively.}
\end{deluxetable*}

\subsection{IRAC Measurements}

In addition to radio emission at $6~\text{cm}$ and $K_{s}$ band emission, \citet{ennis+06} argued that some of the \textit{Spitzer} Infrared Array Camera (IRAC) Bands ($3.6-8~\mu\text{m}$) toward Cas A are dominated by synchrotron emission.  In particular, this study argued that Cas A is morphologically similar in $6~\text{cm}$, IRAC Band 1 (3.6$~\mu\text{m}$), and $K_{s}$ band emission. Therefore, we examine the knots in all four IRAC Bands in addition to radio emission and near-infrared emission. Full analysis (given below) indicates that IRAC Bands 2-4 are dominated by non-synchrotron emission while Band 1 is dominated by synchrotron emission.\footnote{This analysis is based on observations over a four year baseline, spanning from 2001 (date of 6 cm radio imaging) to 2003 (date of near-infrared imaging) to 2005 (date of IRAC imaging).  Synchrotron losses or the injection of fresh electrons into the synchrotron-emitting population may account for small deviations from a reasonable synchrotron spectrum across the entire SED.}

\begin{figure}
	\begin{center}
			\includegraphics[width=0.5\textwidth]{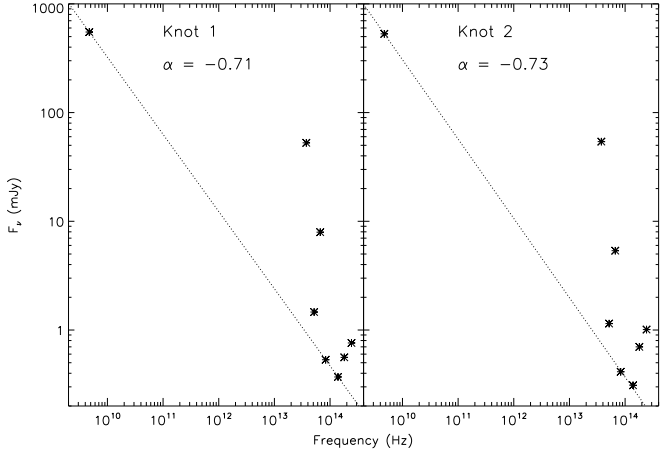}
	\end{center}
	\caption{\scriptsize SED of SEK 1 (left) and SEK 2 (right) including 6 cm radio emission, IRAC Bands 4-1, and $K_{s}$, $H$, and $J$ band emission (indicated by stars from left to right in each panel).  We overplot a synchrotron spectrum (dashed line) corresponding to the flux density of the 6 cm emission and a spectral index of $-0.71$ and $-0.73$ for SEKs 1 and 2, respectively.}\label{fig:SED}
\end{figure}

We plot the SED of knots 1 and 2 across all eight wavebands depicted in \autoref{fig:radio-comp} in \autoref{fig:SED}.  We have fit a synchrotron spectrum to the radio flux densities measured at each knot and assuming a spectral index of $\alpha = 0.78$.  The $K_{s}$ and IRAC Band 1 flux densities are a good match to the assumed synchrotron spectrum, although every other point lies above the inferred power-law.

IRAC Band 2 (4.5$~\mu\text{m}$) is anomalously bright toward the $K_{s}$ band knots with emission an order of magnitude brighter than the inferred synchrotron continuum level in both knots.  \citet{ennis+06} noted this brightness and inferred that line emission must dominate emission toward Cas A between 4 and 5$~\mu\text{m}$.  We suggest the most likely candidate is the CO vibrational band, which has been spectroscopically identified toward Cas A at 4.5$~\mu\text{m}$ \citep{rho+09, rho+12}.  No spectra of the 4.5$~\mu\text{m}$ emission are available toward the southwest shell where these knots occur, although the CO emission observed from AKARI at these wavelengths is generally $\sim60~\text{MJy sr}^{-1}$ ($1.4~\text{mJy arcsec}^{-2}$) \citep{rho+12}, roughly in agreement with the $7.94\pm0.04~\text{mJy}$ we observe over a 1\arcsec.5 radius aperture toward knot 1.

It is known that IRAC Band 4 (8.0$~\mu\text{m}$) is dominated by [Ar\II] (6.99$~\mu\text{m}$) emission \citep[][]{ennis+06,smith+09} while IRAC Band 3 (5.7$~\mu\text{m}$) may be dominated by dust continuum or possibly by forbidden line emission from [Fe\II] (5.3$~\mu\text{m}$).  The IRAC Band 3 emission from both knots is only $\sim1.5$ times the inferred synchrotron continuum level, which may imply a combination of dust continuum and synchrotron emission in Band 3.

Given the good fit between the synchrotron spectrum from $K_{s}$ band and the radio emission, we infer that the knots are both dominated by synchrotron emission in this band.  Henceforth, we refer to these features as synchrotron-emitting knots (SEKs) in $K_{s}$ band.

One intriguing hypothesis regarding the synchrotron emission is that features dominated by synchrotron radiation should be concentrated in compressed magnetic field downstream from the shock front \citep[see, e.g.,][]{reynolds98,bleeker+01}.  It has been argued that this hypothesis is contradicted by the ubiquity of diffuse emission observed in both $K_{s}$ band and radio emission \citep{jones+03}.  However, the presence of compact synchrotron-emitting knots in $K_{s}$ would argue in favor of such a scenario indicating the magnetic field enhancements occur in some locations.

\subsection{Physical Parameters of Synchrotron-Emitting Knots}

Given the multi-epoch imaging of the SEKs in \autoref{fig:knots}, we can use their measured flux densities to estimate the physical conditions required to account for the observed decay rate.

A single electron with Lorentz factor $\gamma$ and emitting synchrotron radiation in a uniform magnetic field with strength $B$ emits power

\begin{equation}
	P = \frac{1}{6\pi}\sigma_{T} c (\gamma^{2} - 1) B^{2}.
\end{equation}

A high-energy ($\gamma \gg 1$) electron with $E = \gamma m_{e} c^{2}$ emitting synchrotron radiation will therefore have a mean lifetime

\begin{equation}
	\tau = \frac{E}{2 \ln (2) P} = \frac{3 \pi m_{e} c}{\ln (2) \sigma_{T} \gamma B^{2}} = \frac{5.6 \times 10^{8}~\mathrm{s}}{\gamma B^{2}}.
\end{equation}

We assume that a SEK will decay in brightness at roughly the same rate.  That is, for a SEK with luminosity $L$ such that $m \sim -2.5 \log (L)$, the magnitude of the synchrotron emission will have a dependence $m \sim -2.5 \log (L_{0} e^{-t/\tau})~\mathrm{mag} \sim 2.5 \log (e) \frac{t}{\tau}~\mathrm{mag} = 0.06 \gamma B^{2} t~\mathrm{mag~yr}^{-1}$. Thus, the magnitude of the synchrotron emission will decay at a rate $\frac{dm}{dt} = 0.06 \gamma B^{2}$ ($B$ in gauss).  The measured decay rate is $0.017~\mathrm{mag~yr}^{-1}$ for SEK 1 across all three epochs and $0.27~\mathrm{mag~yr}^{-1}$ for SEK 2 from the first to second epoch (see note in \autoref{tab:1}). Assuming both SEKs are losing energy in a magnetic field of constant strength and all electrons are emitting as a $\delta$-function at their characteristic frequency $\nu_{c} = \frac{3}{4\pi} \gamma^{2} \frac{e B}{m_{e} c}$ (where $\nu_{c} = 1.4 \times 10^{14}~\mathrm{Hz}$), then the electrons will have $\gamma = 5800 B^{-1/2}$ ($B$ in gauss) and the SEKs will decay at a rate $0.011 B_{\mathrm{mG}}^{3/2}~\mathrm{mag~yr}^{-1}$.  From our measured decay rate, the magnetic fields toward SEK 1 and 2 are approximately $1.3~\mathrm{mG}$ and $5.8~\mathrm{mG}$, respectively.  

To calculate the electron density of each SEK, we make the standard assumption that the number density of relativistic electrons follows a power-law distribution of energies $N(\gamma) = N_{0} \gamma^{-p}$ (where $\gamma \in [1,\infty)$) where the spectral index is given by $\alpha = -(p-1)/2$.  We can integrate this distribution to obtain the total number density of electrons $n_{e} = \int_{1}^{\infty} N_{0} \gamma^{-p} d\gamma = \frac{N_{0}}{p-1}$, such that $N(\gamma) = (p-1) n_{e} \gamma^{-p}$.

The total power per unit volume per unit frequency of emitted synchrotron radiation \citep[see, e.g.,][]{rybickiANDlightman,ghisellini12} from this population of electrons will be

\begin{eqnarray}
	\epsilon_{\nu} & = & \frac{\sigma_{T} c n_{e} B^{2}}{32 \pi^{3/2} \nu_{L}} \left(\frac{\nu}{\nu_{L}}\right)^{-\frac{p}{2}} f(p) \\
	f(p) & \approx & 3^{p/2} (p-1) \left(\frac{2.25}{p^{2.2}} + 0.105\right) \\
	\nu_{L} & = & \left(\frac{e B}{2 \pi m_{e} c}\right)
\end{eqnarray}

\noindent where we have assumed the magnetic field lines in the SEK are tangled, i.e. we have integrated the pitch angle between the path of the electron and the magnetic field over the possible solid angle that synchrotron radiation can be emitted.  At these wavelengths, we assume each SEK is optically thin.  It follows that the total luminosity emitted in near-infrared synchrotron emission at $K_{s}$ band and for each SEK is $L_{K_{s}} = \frac{1}{6} \pi D_{SEK}^{3} \nu \epsilon_{K_{s}}$.  

The observed $K_{s}$ band luminosities of SEK 1 and SEK 2 are $L_{K_{s}} = 7.2 \times 10^{32}~\text{erg s}^{-1}$ and $7.6 \times 10^{32}~\text{erg s}^{-1}$, respectively, defined as $L = 4\pi d^{2} \nu F_{\nu}$ using the flux density from the 2002 epoch and assuming a distance of $d=3.4~\text{kpc}$.  The angular size of both SEKs is approximately $\theta = 3\arcsec$, corresponding to a size of $D_{SEK} = 1.5 \times 10^{17}~\text{cm}$ at $3.4~\text{kpc}$.  For $p = 2.42,2.46$ (corresponding to $\alpha = -0.71,-0.73$ for SEKs 1 and 2, respectively) the electron densities toward SEK 1 and SEK 2 are $15,000~\text{cm}^{-3}$ and $1,000~\text{cm}^{-3}$.

\section{Discussion}

The magnetic field strength in SEKs 1 and 2 is clearly enhanced in a small knot compared to values calculated for acceleration sites distributed across the entire remnant from gamma-ray fluxes.  Moreover, the location of the knots places them approximately $123\arcsec$ from the expansion center of the remnant given their location in \autoref{fig:knots} and the expansion center calculated in \citet{thorstensen+01}.  It is therefore reasonable to assume that these knots are behind the location of the forward shock at this position angle \citep[e.g., in][]{fesen+01,gotthelf+01}.  

The presence of compact knots of synchrotron emission closely matches the emission model derived in \citet{reynolds98} where the magnetic field can be compressed and amplified downstream of the forward shock.  For a compression ratio $r=4$ (i.e., $B_{1} = r B_{2}$), which is typical for the adiabatic strong-shock limit with $\gamma=5/3$, the magnetic field strengths calculated for SEKs 1 and 2 ($B_{1}$) imply an upstream magnetic field strength ($B_{2}$) of $0.33-1.5~\text{mG}$, which is of the same order, but somewhat larger than the values inferred from gamma-ray measurements. This rough agreement supports the overall validity of the \citet{reynolds98} emission model. 

Additional observations and analysis are needed to determine if the field values above the predictions can be explained within this model. Otherwise, some additional mechanism may be amplifying the magnetic field in this region, such as turbulent amplification brought on by Rayleigh-Taylor instabilities in the shock \citep[as in, e.g.,][]{jj99}. 

Moreover, it is curious that SEK 2 is fading so quickly (\autoref{fig:knots}) and that this behavior implies a large magnetic field strength and low electron density compared to SEK 1.  There must be a sharp change in the properties of the remnant over a $\sim0.05~\text{pc}$ distance such that, despite being entrained in the other synchrotron-emitting ejecta, the SEKs have their own distinct properties.  Along with the additional magnetic field amplification, this behavior also supports a scenario in which a strong shock interaction and instabilities in the ejecta have compressed knots of material.  Indeed, the presence of a recent shock interaction to the southwest of Cas A would also explain enhancement in hard X-ray and radio emission \citep{anderson+95,grefenstette+15} as well as the presence of shocked molecular gas \citep{kilpatrick+14} toward this part of the remnant.

\section{Conclusions}

We have presented multi-epoch $K_{s}$ band imaging of Cas A demonstrating that two knots of emission to the southwest of the remnant appear entrained in the ejecta.  Using radio and \textit{Spitzer} IRAC data from the literature, we have argued that the $K_{s}$ band emission from these knots is dominated by synchrotron radiation.  Our main conclusions from our analysis of these knots are:

\begin{enumerate}
	\item The two synchrotron-emitting knots are fast moving ($\sim5000~\text{km s}^{-1}$) and appear to have magnetic fields of $1.3$ and $5.8~\text{mG}$.
	\item These fields are an order of magnitude higher than deduced by gamma ray measurements that average electron acceleration sites distributed over the entire remnant.
	\item This behavior appears to be consistent with the compression and amplification of the magnetic field downstream of the forward shock as proposed by \citet{reynolds98}.
\end{enumerate} 

\acknowledgments

We thank Jeonghee Rho for providing $K_{s}$ band imaging for this publication.  Funding for this research was provided by NASA through Contract Number 1255094 issued by JPL/Caltech.  We would also like to acknowledge the PISCES instrument team, especially Don McCarthy and Craig Kulesa, for their assistance in performing the new $K_{s}$ band observations presented herein.


\bibliography{masterbib}

\clearpage



\end{document}